\documentclass[11pt]{article}

\usepackage{verbatim}
\usepackage{amsmath}
\usepackage{amsfonts}
\usepackage{amssymb}
\usepackage{latexsym}
\usepackage{graphics}
\usepackage{lscape} 
\usepackage{epsfig}
\usepackage{color}
\usepackage{cite}

\newcommand{\tbl}{
\begin{table}
\[
\arry{l | c | c | c | c }{
& \text{Vol} & \text{Vol}_K & \gL_W & \gL_K 
\\[0.5ex]\hline&&&&\\[-2.2ex] 
S^1 & R & \frac 1{\dsp R} & 2\gp\, R\, \Intr & \frac 1{\dsp R}\, \Intr 
\\[.5ex]\hline&&&&\\[-2.2ex] 
T^2 & R_5 R_6\,\sin \gvth &  \frac1{\dsp R_5 R_6\, \sin \gvth} & 
\gp\, \big( R_5\, \Intr + e^{i\gvth} R_6\, \Intr \big) & 
\frac i{\dsp \sin \gvth} 
\Big( \frac {\dsp e^{-i\gvth}}{\dsp R_5}\, \Intr + \frac 1{\dsp R_6} \, \Intr   \Big) 
}
\]
\caption{This table summarizes our notation for the circle and  the
torus: $\text{Vol}_W = (2\gp)^{D-d} \text{Vol}$ with 
$D-d=1,2$, respectively, and $\text{Vol} \cdot \text{Vol}_K = 1$. 
}
\labl{tb:VolLatt} 
\end{table}
}

\renewcommand{\d}{\mathrm{d}}

\DeclareMathSymbol{\mg}{\mathrel}{symbols}{"1D}

\newcommand{\ga}{\alpha}
\newcommand{\gb}{\beta}
\renewcommand{\gg}{\gamma}
\newcommand{\gd}{\delta}
\renewcommand{\ge}{\epsilon}

\newcommand{\gf}{\phi}

\newcommand{\gm}{\mu}
\newcommand{\gn}{\nu}

\newcommand{\gr}{\rho}
\newcommand{\gth}{\theta}
\newcommand{\gvth}{\vartheta}

\newcommand{\gt}{\tau}

\newcommand{\gp}{\pi}
\newcommand{\gps}{\psi}
\newcommand{\get}{\eta}

\newcommand{\gG}{\Gamma}
\newcommand{\gD}{\Delta}
\newcommand{\gF}{\Phi}

\newcommand{\gL}{\Lambda}
\newcommand{\gS}{\Sigma}

\newcommand{\cL}{{\cal L}}

\newcommand{\cN}{{\cal N}}

\newcommand{\cP}{{\cal P}}

\newcommand{\tA}{{\tilde A}}

\newcommand{\tJ}{{\tilde J}}

\newcommand{\tr}{\text{tr}}

\newcommand{\Slashed}{\hspace{-1.4ex}/\hspace{.2ex}}

\newcommand{\ra}{\rightarrow}

\newcommand{\der}{\partial}

\newcommand{\inv}{^{-1}}
\newcommand{\dsp}{\displaystyle}

\newcommand{\labl}[1]{\label{#1}}
\newcommand{\beq}{\begin{equation}}
\newcommand{\eeq}{\end{equation}}
\newcommand{\barr}{\begin{array}}
\newcommand{\earr}{\end{array}}
\newcommand{\equ}[1]{\begin{gather} #1 \end{gather}}
\newcommand{\equa}[1]{\begin{align} #1 \end{align}}
\newcommand{\items}[1]{\begin{itemize} #1 \end{itemize}}

\newcommand{\arry}[2]{\begin{array}{#1} #2 \end{array}}

\newcommand{\pmtrx}[1]{\begin{pmatrix} #1 \end{pmatrix}}

\newcommand{\non}{\nonumber}
\newcounter{oldcounter}

\newcommand{\bder}{\bar\partial}
\newcommand{\bn}{{\bar n}}

\newcommand{\bw}{{\bar w}}

\newcommand{\bz}{{\bar z}}

\newcommand{\bD}{{\bar D}}

\newcommand{\bS}{{\bar S}}

\newcommand{\bga}{{\bar \alpha}}
\newcommand{\bgb}{{\bar\beta}}

\newcommand{\bgF}{{\bar\Phi}}

\newcommand{\tgd}{{\tilde \delta}}

\newcommand{\tgf}{{\tilde \phi}}

\newcommand{\Intr}{\mathbb{Z}}
\newcommand{\Cplx}{\mathbb{C}}

\newcommand{\ba}[2]{\[\begin{array}{#2}\label{#1}}
\newcommand{\ea}{\end{array}\]}
\newcommand{\be}{\begin{equation}}
\newcommand{\ee}{\end{equation}}
\newcommand{\bea}{\begin{eqnarray}}
\newcommand{\eea}{\end{eqnarray}}

\newcommand{\U}[1]{\mathrm{U(#1)}}

\newcommand{\brkt}[2]{\bigl[ ^{#1}_{#2} \bigr]}

\newcommand{\sm}{{\,\mbox{-}}}

\newcommand{\sfrac}[2]{\mbox{$\frac{#1}{#2}$}}

\begin{document}

\thispagestyle{empty}

\begin{flushright}
FTPI-MINN-05/06 \\ 
UMN-TH-2347/05  \\      
hep-th/0503153
\end{flushright}
\vskip 2 cm
\begin{center}
{\Large {\bf 
Renormalization of Supersymmetric Gauge Theories on Orbifolds: 
\\[2ex] 
Brane Gauge Couplings and Higher Derivative Operators 
}
}
\\[0pt]
\vspace{1.23cm}
{\large
{\bf Stefan Groot Nibbelink$^{a,}$\footnote{
{{ {\ {\ {\ E-mail: nibbelin@hep.umn.edu}}}}}}},
{{\bf Mark Hillenbach$^{b,}$\footnote{
{{ {\ {\ {\ E-mail: mark@th.physik.uni-bonn.de}}}}}}} 
\bigskip }\\[0pt]
\vspace{0.23cm}
${}^a$ {\it 
William I. Fine Theoretical Physics Institute,
School of Physics \& Astronomy, \\
University of Minnesota, 116 Church Street S.E., 
Minneapolis, MN 55455, USA 
\\[1ex]}${}^b$ {\it 
Physikalisches Institut, Universit\"at Bonn, 
Nussallee 12, D-53115 Bonn, Germany
\\}
}
\bigskip
\vspace{1.4cm} 
\end{center}
\subsection*{\centering Abstract}

We consider supersymmetric gauge theories coupled to hyper multiplets
on five and six dimensional orbifolds and determine the bulk and local
fixed point renormalizations of the gauge couplings. We infer from a
component analysis that the hyper multiplet does not induce 
renormalization of the brane gauge couplings on the five dimensional
orbifold $S^1/\Intr_2$. This is not due to supersymmetry, since the
bosonic and fermionic contributions cancel separately. We extend this
investigation to $T^2/\Intr_N$ orbifolds using supergraph techniques
in six dimensions. On general $\Intr_N$ orbifolds the gauge couplings
do renormalize at the fixed points, except for the $\Intr_2$ fixed
points of an even ordered orbifold. To cancel the bulk one--loop
divergences a dimension six higher derivative operator is needed, in
addition to the standard bulk gauge kinetic term.

\newpage

\setcounter{page}{1}

\section{Introduction and summary}

The investigation of theories of extra dimensions has been an active
field of research initiated by
\cite{Arkani-Hamed:1998rs,Antoniadis:1998ig}.  
Most of the phenomenological activity has focused
on five dimensional (5D) models, in particular models on simple
orbifolds like $S^1/\Intr_2$ or $S^1/\Intr_2\times \Intr_2'$
\cite{Mirabelli:1997aj,Barbieri:2000vh,Delgado:1998qr}.  
An important issue of such investigations was the running of the 4D 
gauge coupling in extra dimensions and possible gauge
coupling unification \cite{Dienes:1998vh,Dienes:1998vg}. 
A complication is that the gauge couplings are sensitive to the 
ultra--violet (UV)
completion of the theory \cite{Hebecker:2002vm}. In this Letter we 
study the gauge coupling running by calculating the self--energy in
extra dimensions. In particular, we investigate the renormalization of
bulk and fixed point gauge operators in supersymmetric (SUSY) field
theories on 5D and 6D orbifolds.

As a warm up, we start our analysis with a single complex scalar
coupled to a gauge field in the bulk of $S^1/\Intr_2$. To cancel the
divergences of  
the scalar loop both bulk and brane localized counter terms are needed
for the gauge field. This result is an example of the generic fact that
on an orbifold both bulk and fixed point localized operators
renormalize \cite{Georgi:2000ks,vonGersdorff:2003dt,GrootNibbelink:2003gd}. 
However, such localized counter terms are not always required: 
A charged bulk fermion does not require counter terms
for the gauge field at the orbifold fixed points. The absence of
brane gauge counter terms persists in SUSY models, because the
contributions of the complex scalars of the hyper multiplet also cancel.

This raises the question, whether this is an accident of the 
simple $S^1/\Intr_2$ orbifold or holds more generically for
$T^2/\Intr_N$ orbifolds in 6D SUSY theories. We investigate this
question by computing the one--loop self--energy for the vector
multiplet in 6D.  To this end we set up a 6D extension of $\cN=1$
supergraphs based on representing 6D SUSY theories by 
$\cN=1$ 4D superfields
\cite{Arkani-Hamed:2001tb,Hebecker:2001ke,Marti:2001iw}.  
We find that for generic $\Intr_N$ orbifolds the gauge couplings
at almost all fixed points do renormalize due to bulk hyper
multiplets. There is no contradiction with the 5D 
$S^1/\Intr_2$ result, because $\Intr_2$ fixed points of even
ordered orbifolds (and therefore $\Intr_2$ orbifolds in particular) are
the only fixed points that do not receive any gauge coupling
renormalization.

Since we compute the full one--loop gauge multiplet self--energy, we
can determine the bulk renormalization of the gauge multiplet. 
We find that a dimension six higher derivative term for the gauge
multiplet is generated.  (Higher derivative counter terms are also needed
in 5D orbifold models if brane localized interactions
for bulk fields are considered \cite{Ghilencea:2004sq}.) Such higher derivative
theories may have remarkable UV properties \cite{Smilga:2004cy}: The
higher derivative operators act as regulators that make many loop
graphs finite. Higher derivative hyper multiplet operators do
not seem to be allowed by gauge and 6D Lorentz 
invariance combined. (All gauge coupling corrections at one loop would
be finite if they were present.)

Let us close with a few comments on the context and possible
extensions of our work. In 6D the constraints of anomalies
are very severe \cite{Seiberg:1996qx,Danielsson:1997kt}, but since we
were only interested in the gauge coupling running, we do
not take these constraints into account. Moreover, we restrict
ourselves to Abelian theories only; in a future publication \cite{GNH}
we investigate non--Abelian theories and work out the details of the
threshold corrections we identify.   
Our investigation is restricted to one--loop corrections only. 
However, we expect that the results in fact hold to all orders in
perturbation theory up to infra--red (IR) effects. Both at the fixed
points and in the bulk holomorphicity arguments 
\cite{Shifman:1986zi,Shifman:1991dz,Seiberg:1993vc,Seiberg:1994bp,Weinberg:2000cr}
of $\cN=1$ SUSY field theories in 4D apply.

The outline is as follows: In section \ref{sc:S1} we
study the running of local gauge couplings due to scalars and fermions
on $S^1/\Intr_2$. In section \ref{sc:Susy6D} we perform a manifestly
SUSY one--loop computation of the gauge multiplet
self--energy on generic $T^2/\Intr_N$ in 6D. We determine the bulk and
fixed point renormalizations of the gauge coupling and identify a
higher derivative operator in the bulk. In the Appendix we describe
the regularization of the divergent integral encountered in 4, 5 and 6D.

\section*{Acknowledgments}

We would like to thank 
R.\ Auzzi, T.\ Gherghetta, D.\ Ghilencea, A.\ Hebecker, H.\ P.\ Nilles, M.\ Shifman, M.\
Trapletti, A.\ Vainshtein and M.\ Voloshin for simulating and 
useful discussions. SGN would like to thank the Bonn University for
their kind hospitality at various stages of this project. MH thanks
the University of Minnesota, also for kind hospitality. 
The work of
SGN has been supported in part by the Department of Energy 
under contract DE--FG02--94ER40823 at the University of Minnesota.
The work of MH was partially supported by the EU 6th Framework
Program MRTN-CT-2004-503369 ``Quest for Unification'' and
MRTN-CT-2004-005104 ``Forces Universe''.

\section{Bulk and fixed point localized corrections on
$\boldsymbol{S^1/\Intr_2}$} 
\labl{sc:S1}

\subsection{Scalar on $\boldsymbol{S^1/Z_2}$}
\labl{sc:scalarS1}

We begin our analysis with a complex scalar $\tgf$
coupled to a $\U{1}$ gauge field $\tA_M$ in 5D 
compactified on $S^1/\Intr_2$. The coordinate $y$ of the covering
circle $S^1$ is periodic $y \sim y + 2\pi\, R$. The $\Intr_2$
reflection acts on these fields as 
\equ{
\tgf(-y) ~=~ Z \, \tgf(y)~, 
\qquad 
\tA_\gm(-y) ~=~ \tA_\gm(y)~, 
\qquad 
\tA_5(-y) ~=~ - \tA_5(y)~, 
\labl{S1cond} 
} 
where we have suppressed the 4D coordinate $x^\gm$. 
To be able to trace the dependence on the orbifold boundary
conditions, we keep the parity eigenvalue $Z = \pm$ of the scalar
$\tgf$ arbitrary. In many studies of the orbifold $S^1/\Intr_2$ the
fields are expanded into even and odd mode functions. For sufficiently
simple orbifolds this is a useful procedure, but since we want to
extend our analysis eventually to more complicated orbifolds, we
choose instead to obtain orbifold compatible fields from fields
defined on the covering space \cite{Georgi:2000ks}. 
For example, let $\gf$ be a complex scalar on the covering circle. By
employing an orbifold projector we obtain a field $\tgf$ satisfying
\eqref{S1cond} as  
\equ{
\tgf(y) ~=~ \frac 12 \Big( \gf(y) + Z \, \gf(-y) \Big)~, 
\labl{compatible}
}
where the extensions are obvious. 
We define orbifold compatible functional differentiation as 
\equ{
\tgd_{21} ~=~ \frac {\gd \, \tJ_2}{\gd\, \tJ_1} ~=~ 
\frac 12 
\Big( \gd^5(y_2 - y_1) + Z\, \gd^5(y_2+y_1)\Big)~, 
\labl{OrbiDelta}
}
where $\tJ$ is the source coupled to $\tgf$. Here and throughout the
paper we only indicate the internal
coordinate(s) explicitly where the orbifolding is non--trivial, i.e.\
$\gd^5(y_2 \pm y_1) = \gd^4(x_2-x_1) \gd(y_2\pm y_1)$.

We used this method to obtain the gauge field self--energy at one loop
due to the complex scalar $\tgf$ with charge $q$.  
There is a tadpole (seagull) diagram: 
\equ{
\raisebox{-.5ex}{\scalebox{0.3}{\mbox{\begin{picture}(0,0)%
\includegraphics{AA_T.pstex}%
\end{picture}%
\setlength{\unitlength}{4144sp}%
\begingroup\makeatletter\ifx\SetFigFont\undefined%
\gdef\SetFigFont#1#2#3#4#5{%
  \reset@font\fontsize{#1}{#2pt}%
  \fontfamily{#3}\fontseries{#4}\fontshape{#5}%
  \selectfont}%
\fi\endgroup%
\begin{picture}(2519,1409)(2004,-5042)
\end{picture}
}}}
~=~ q^2\, \int (\d^5 X)_{12} \, \tA_1^M \tA_1^N\, \get_{MN}\,  
\tgd_{21} \frac{1}{(\Box_5 - m^2)_2} \tgd_{21}~, 
}
and a genuine self--energy diagram 
\equa{\dsp 
\raisebox{-1.4ex}{\scalebox{0.3}{\mbox{\begin{picture}(0,0)%
\includegraphics{AA_S.pstex}%
\end{picture}%
\setlength{\unitlength}{4144sp}%
\begingroup\makeatletter\ifx\SetFigFont\undefined%
\gdef\SetFigFont#1#2#3#4#5{%
  \reset@font\fontsize{#1}{#2pt}%
  \fontfamily{#3}\fontseries{#4}\fontshape{#5}%
  \selectfont}%
\fi\endgroup%
\begin{picture}(4544,928)(1554,-3450)
\end{picture}
}}}
~=~  q^2\, \int &(\d^5 X)_{12} \, \tA_1^M \tA_2^N\, 
\Big( 
\frac{1}{(\Box_5 - m^2)_2} \tgd_{21} \,
\frac{\der_{1\, M} \der_{2\,N}}{(\Box_5 - m^2)_2}
 \tgd_{21}  
~+~ \non \\[2ex]\dsp  &
~-~ \frac{\der_{1\,M}}{(\Box_5 - m^2)_2} \tgd_{21}\, 
\frac{\der_{2\, N}}{(\Box_5 - m^2)_2} \tgd_{21} 
\Big)~. 
}
Here $(\d^5 X)_{12}$ denotes the integration over the coordinates 
$X_1^M = (x_1^\gm, y_1)$ and $X_2^M = (x_2^\gm, y_2)$, and 
partial differentiation w.r.t. $X_2^M$ is indicated by 
$\der_{2\,M} = \der/\der X_2^M$. The spacetime metric $\get_{MN}$ 
uses the mostly plus convention, and the 4D and 5D 
kinetic operators read $\Box = \der^\gm \der_\gm$ and 
$\Box_5 = \Box + \der_5^2$, respectively. Notice that all terms in
both expressions contain two orbifold 
delta functions $\tgd_{21}$, i.e.\ the orbifold projector is inserted
twice. Since a projector squared is the projector again, one of them
can be replaced by a conventional delta function 
$\tgd_{21} \ra \gd_{21} = \gd^4(x_2-x_1)\gd(y_2-y_1)$. 
This can be confirmed explicitly by inserting \eqref{OrbiDelta} for
one of the orbifold delta functions and perform a change of coordinates
$y_2 \ra - y_2$. The leftover $\tgd_{21}$ consists of two parts, see
\eqref{OrbiDelta}: The first part, $\frac 12 \, \gd_{21}$, gives
rise to contributions in 5D 
compactified on a circle, with an additional normalization factor of
$\frac 12$. The second part of the orbifold delta function reads
$\frac 12 Z\, \gd^4(x_2-x_1) \gd(y_2 + y_1)$. If there were no
derivatives, integration over $y_2$ would lead to the fixed point delta
function $\gd(2y_1)$ and hence to localization at the orbifold fixed
points. In the presence of the $y$ derivatives in the propagators
the amplitude acquires non--local contributions which are sourced
by the fixed points. However, the counter terms needed to cancel the 
divergences are local. As the 4D--localized parts
are proportional to the factor $Z$, it follows that for two complex
scalars of opposite parities (and equal or opposite charges) all
localized contributions cancel identically.

\subsection{Fermion on $\boldsymbol{S^1/\Intr_2}$}
\labl{sc:fermionS1}

Next we move to a Dirac fermion $\gps$ on the same orbifold, which
satisfies the boundary conditions 
\equ{
\widetilde{\gps}(-y) ~=~ \gg_5\, \widetilde{\gps}(y)~, 
\qquad 
\widetilde{\overline{\gps}}(-y) ~=~ \widetilde{\overline{\gps}}(y)\, (-\gg_5)~, 
}
so that the kinetic terms are invariant. The functional derivative 
w.r.t.\ the source $\tJ$ for $\widetilde{\overline{\gps}}$ reads 
\equ{
\tgd_{21} ~=~ \frac {\gd \, \tJ_2}{\gd\, \tJ_1} ~=~ 
 \frac 12 
\Big( \gd^5(y_2 - y_1) - \gg_5\, \gd^5(y_2+y_1)\Big)~, 
\labl{OrbiDeltaF}
} 
where we again suppressed the 4D coordinate dependence in the delta
function. 
Functional differentiation w.r.t.\ the source $\tilde{\bar{J}}$ for $\widetilde{\gps}$
defines $\tilde{\bar{\gd}}$ in a similar fashion; it is obtained from $\tgd$
by replacing $\gg_5 \ra - \gg_5$. Using similar steps as in the 
scalar calculation, we can evaluate the photon self--energy due to the
fermion: All orbifold projectors in the loop can be removed except for
one  
\equ{
\raisebox{-1.4ex}{\scalebox{0.3}{\mbox{\begin{picture}(0,0)%
\includegraphics{AA_F.pstex}%
\end{picture}%
\setlength{\unitlength}{4144sp}%
\begingroup\makeatletter\ifx\SetFigFont\undefined%
\gdef\SetFigFont#1#2#3#4#5{%
  \reset@font\fontsize{#1}{#2pt}%
  \fontfamily{#3}\fontseries{#4}\fontshape{#5}%
  \selectfont}%
\fi\endgroup%
\begin{picture}(4544,1154)(1554,-3563)
\end{picture}
}}}
~=~ 
\gS_F ~=~ \frac {q^2}2 \,\int(\d X)_{12} \, 
\tr\Big[  A\Slashed_1 (\der\Slashed+m)\inv_1\gd_{12} 
A\Slashed_2(\der\Slashed+m)_2\inv \tgd_{21} \Big]~. 
}
Here the trace is over the four component spinor indices and 
$A\Slashed = A^M \gg_M$. 
Again we see from the expression for the delta function for the fermion
\eqref{OrbiDeltaF} that the amplitude consists of 5D and 4D localized
parts. In fact, the localized part vanishes:

We expand the localized part in momentum space: 
\equa{
\gS_{F\,4D} ~=~ &  - \frac {q^2}4 \int \frac{\d^4 p\, \d^4 k}{(2\pi)^{8}} 
\frac 1{(2\pi\, R)^2} \sum_{n_1,n_2 \in \Intr/R} 
\frac 1{[p^2 +n_3^2 + m^2] [(p+k)^2 +n_4^2 +m^2]} \, \cdot 
\non \\[1ex] 
& \cdot \bigg\{ 
A^\gm(k,n_1) A^\gn(-k,n_2) \, 
\tr\Big[\gg_5 \, \gg_\gm  \, (p\Slashed +n_3 \gg_5 + i m) \, 
\gg_\gn\, (p\Slashed + k\Slashed +n_4 \gg_5 + i m) \Big]
\\[1ex]
& \quad + A^5(k,n_1) A^5(-k,n_2) \, 
\tr\Big[\gg_5^2 \,  (p\Slashed +n_3 \gg_5 + i m) \, 
\gg_5 \, (p\Slashed + k\Slashed +n_4 \gg_5 + i m) \Big]
\bigg\}~.
\non 
}
Here $p,k$ are 4D (loop) momenta. The loop 
Kaluza--Klein (KK) momenta $2n_3 = n_2-n_1$ and 
$2n_4 = -n_2-n_1$ are expressed in terms of those of the external
photons. As these are localized contributions, the KK number is not
preserved: $n_2$ need not be equal to $-n_1$. Instead, 
$n_1$ and $n_2$ are either both even or both odd, hence there is no
mixing between $A^5$ and $A^\mu$. The presence of $\gg_5$ in these
expressions shows that all traces vanish identically except for 
$\tr[\gg_5\gg_\gm p\Slashed \gg_\gn k\Slashed]$. By employing a
Feynman parameterization of the propagators, the loop integral 
implies that $p_\gr \sim k_\gr$, and therefore also this trace
vanishes.

\subsection{Hyper multiplet gauge coupling renormalization on
$\boldsymbol{S^1/\Intr_2}$}
\labl{sc:HyperS1}

We use the previous results to get some
feeling for the localization of gauge couplings in SUSY 
theories: The two chiral multiplets inside a hyper multiplet have
opposite $\Intr_2$ boundary conditions. From section 
\ref{sc:scalarS1} we know that two scalars with opposite boundary
conditions do not give localized gauge 
coupling contributions. And in section \ref{sc:fermionS1} we reached
the same conclusion for a Dirac fermion, i.e.\ two chiral fermions
with opposite charges and boundary conditions. This implies that the
hyper multiplet will not lead to any localized gauge coupling
renormalization.

We have confirmed that no brane localized gauge counter terms are
needed by performing an explicit supergraph calculation of the 
$VV$--, $\bS S$-- and $S V$--selfenergies that are given in figure
\ref{fg:VV_VS_bSS}. (Details will be presented in the next section in
6D.) The 5D bulk gauge coupling renormalizes
as 
\equ{
\frac 1{g_R^2} ~=~ \frac 1{g^2} - \frac {2 q^2}{(4\gp)^2}\,  |m|~, 
}
where the subscript $R$ refers to the renormalized gauge coupling. 
This result is compatible with the results obtained by Witten
\cite{Witten:1996qb} and used by Seiberg and others
\cite{Seiberg:1996bd,Morrison:1996xf,Intriligator:1997pq} to analyze
SUSY gauge theory in non--compact 5D.

\section{Supersymmetric gauge theories with matter on
6D orbifolds}
\labl{sc:Susy6D}

We investigate SUSY theories on an arbitrary 6D 
orbifold $T^2/\Intr_N$. (In fact, $N$ is crystallographically constrained
to be either $2,3,4,6,7,8 \text{ or } 12$.) The field content we
consider is a charged hyper multiplet coupled to a gauge multiplet. 
We employ $\cN=1$ 4D superfields to describe these multiplets 
\cite{Marcus:1983wb,Arkani-Hamed:2001tb,Hebecker:2001ke,Dudas:2004ni},
and the superspace conventions of Wess \& Bagger 
\cite{Wess:1992cp}. 
The gauge multiplet contains a vector multiplet $V$ and a
chiral multiplet $S$. The hyper multiplet consists of two chiral
multiplets $\gF_\pm$ that are charged oppositely. The superfields are
made orbifold compatible using methods similar to
\eqref{compatible}. In order to keep the notation simple, we have
dropped the twiddles on them.

We employ complex coordinates $z = \frac 12(x_5 - i x_6)$ and 
$\bz = \frac 12(x_5 + i x_6)$, so that we find for the derivatives: 
$\der = \der_5 + i \der_6$, $\bder = \der_5 - i \der_6$ and 
$\der \bder = \der_5^2 + \der_6^2$. (The reduction to 5D is
straightforward: Set  
$z = \bz = \frac 12 y$, $R_5=R$ and $\der = \bder = \der_5$.) 
The periodicity conditions of the torus $T^2$, 
$z \sim z+ \gp\, R_1 \sim z+ \gp\, e^{i\gvth} R_2$, 
define the ``winding mode'' lattice $\gL_W$. This lattice, the
KK lattice $\gL_K$ and the volumes of their fundamental
domains are collected in table \ref{tb:VolLatt}. 
An orbifold $T^2/\Intr_N$ is obtained by requiring that the field
theory on the covering torus $T^2$ is invariant under the 
$\Intr_N$ rotation: 
\equ{
z \ra e^{-i\, \gf} \, z~, 
\quad 
\bz \ra e^{i\, \gf}\, \bz~,
\qquad 
\der \ra e^{i\,\gf}\, \der~, 
\quad 
\bder \ra e^{-i\, \gf}\, \bder~, 
}
where the phase $\gf$ is such that $e^{i\, N\gf} = 1$. In order for
this $\Intr_N$ orbifold action to be compatible with the lattice,
conditions on the radii $R_5$, $R_6$ and phase $\gvth$ may apply. 
(For example for a $\Intr_3$ orbifold $R_5=R_6=R$ and 
$\gvth=\gf=2\gp/3$.) The superfields $V, S, \gF_+$ and $\gF_-$
transform as  
\equ{
V \ra V, 
\quad 
S \ra e^{i\, \gf}\, S, 
\qquad 
\gF_\pm \ra e^{i \, a_\pm \gf}\, \gF_\pm~. 
}
Only for the hyper multiplet we have an arbitrary integer 
$0 \leq a_+ \leq N-1$ since $a_- = N -1 - a_+$. (Note that this is
compatible with the $\Intr_2$ case: There one chiral multiplet is
even and the other is odd.) As in section \ref{sc:scalarS1}, we 
define the orbifold delta function as 
\equ{
\tgd^{(a)}_{21} ~=~ 
\frac 1N \, \sum_{b=0}^{N-1}\, e^{i\, ba \, \gf}\, 
\gd \big( z_2 - e^{i\, b\gf} z_1 \big)~, 
\labl{T2ZNdelta}
}
where 
\( 
\gd(z_2 - z_1) = \gd^2(z_2- z_1)\, \gd^4(x_2-x_1) \,
\gd^4(\gth_2-\gth_1), 
\)  
for a superfield that transforms with a phase $e^{i\,a\gf}$. With this
formalism we can set up a supergraph formalism 
\cite{Wess:1992cp,West:1990tg,Gates:1983nr} for
orbifold theories.

\tbl

We close this introductory section with an exposition of the
relevant Lagrangians written in terms of $\cN=1$ superfields. 
The gauge invariant bulk vector multiplet Lagrangian can be written 
as
\equ{
\cL_{\rm gauge} ~=~ 
\frac 1{2g^2N} 
\int \d^2 \gth\, W^\ga W_\ga \,+\,  
\frac 1{g^2N} 
\int \d^4 \gth\, \Big(
\der V \bder V +  \bS S  - \sqrt 2\, \bder V S  - \sqrt 2\, \der V \bS 
\Big)~, 
\labl{Lgauge}
} 
where $W^\ga = - \frac 14 \bD^2 D_\ga V$ is the 4D 
superfield strength, and $1/g^2$ is the mass dimension two gauge
coupling.  The factor $1/N$ in the Lagrangian \eqref{Lgauge} is
included, because we perform all our calculations on the covering
space of the 
$T^2/\Intr_N$ orbifold. In addition, for orbifolds we can have fixed
point localized 4D gauge actions of the form 
\equ{
\cL_{\rm gauge}^{\rm fix} ~=~ 
\sum_{b=1}^{N-1}\, \frac 1{2g^2_b N} 
\int \d^2 \gth\, W^\ga W_\ga 
\, 
\gd^2\big( ( 1 - e^{i\, b \gf} ) z\big)~, 
\labl{LgaugeFix} 
} 
where the gauge couplings $1/g_b^2$ are dimensionless.
Note that they have a non--standard normalization and that $g_{N-b} =
g_b$. The gauge invariant Lagrangian for the hyper multiplet with
charge $q$ reads  
\equ{
\cL_{\rm hyper} ~=~ \frac 1{N} \int \d^2 \gth\,
 \gF_- \Big( \der + \sqrt 2 q\, S \Big)  \gF_+ 
+ \text{h.c.} + \frac 1{N} 
\int \d^4 \gth\, \bgF_\pm e^{\pm 2 q\, V} \gF_\pm~, 
\labl{Lhyper}
} 
where in the last term summation over $+$ and $-$ is implied. 
The Hermitian conjugation acts on the chiral superfields 
as well as on the holomorphic derivative $\der$. In 6D the hyper
multiplet is massless \cite{Sierra:1983fj,Howe:1983fr}, while in 5D it
can have a real mass  
$m(y) = m \, \ge(y)$, with $\ge(y)$ the step function on $S^1$, which
can be thought of as the vacuum expectation value of the real part of
$S$.

\subsection{Bulk and fixed point localized gauge selfenergies on 
$\boldsymbol{T^2/\Intr_N}$}
\labl{sc:hyperT2}

We investigate the renormalization of (localized) gauge couplings on 
6D orbifolds. As we consider an Abelian theory, only the hyper
multiplet loops lead to gauge coupling renormalization. 
The propagators of chiral components of the hyper multiplet read 
\equ{
\bgF_\pm 
\raisebox{-.2ex}{\scalebox{0.4}{\mbox{\begin{picture}(0,0)%
\includegraphics{bPhPh_prop.pstex}%
\end{picture}%
\setlength{\unitlength}{4144sp}%
\begingroup\makeatletter\ifx\SetFigFont\undefined%
\gdef\SetFigFont#1#2#3#4#5{%
  \reset@font\fontsize{#1}{#2pt}%
  \fontfamily{#3}\fontseries{#4}\fontshape{#5}%
  \selectfont}%
\fi\endgroup%
\begin{picture}(2744,254)(2679,-4688)
\end{picture}
}}}
\gF_\pm 
~=~ 
\frac 1{\Box_6}~, 
\qquad 
\gF_+
\raisebox{-.2ex}{\scalebox{0.4}{\mbox{\begin{picture}(0,0)%
\includegraphics{PhPh_prop.pstex}%
\end{picture}%
\setlength{\unitlength}{4144sp}%
\begingroup\makeatletter\ifx\SetFigFont\undefined%
\gdef\SetFigFont#1#2#3#4#5{%
  \reset@font\fontsize{#1}{#2pt}%
  \fontfamily{#3}\fontseries{#4}\fontshape{#5}%
  \selectfont}%
\fi\endgroup%
\begin{picture}(2744,254)(2679,-4688)
\end{picture}
}}}
\gF_-
~=~ 
\frac {\bder}{\Box_6}\, \frac {D^2}{\sm 4 \Box} ~, 
\labl{Props}
} 
where $\Box_6 = \Box + \der \bder$. (In 5D these propagators may
contain the mass $m$ of the hyper multiplet.)
We computed the one--loop self energy diagrams for external
superfields $VV$, $VS$ and $\bS S$, given in figure
\ref{fg:VV_VS_bSS}, on the $T^2/\Intr_N$ orbifold. The tadpole graph
cancels gauge non--invariant contributions from the other two graphs of
$\gS_{VV}$. By including the superfields $V$, $S$ and $\bS$
in the amplitudes, the sum of the supergraphs
$\gS = \gS_{VV}+\gS_{V S}+\gS_{V \bS}+\gS_{\bS S}$ 
becomes  
\equ{
\gS ~=~ \frac {2q^2}{N}\, \sum_{b=0}^{N-1} \, 
\int (\d^6 X)_{12} \d^4 \gth \, 
\cP_{b} (X_2, X_1)\,
\cos \big(a_+ + \sfrac 12\big) b \gf \, 
\Big\{
\, \cos \big( \sfrac 12 b \gf\big) \, 
V_2 \frac {(D^\ga \bD^2 D_\ga)_1}8 V_1 ~+~
\non \\[0ex]
~+~
 \bder_2 V_2 \der_1 V_1 
~+~ 
 \bS_2 S_1
~-~ 
\sqrt 2 \, \bder_2 V_2 S_1
~-~ 
\sqrt 2 \, \bS_2 \der_1 V_1
\, 
\Big\}~. 
\labl{1LoopAmpl}
} 
We have replaced the two orbifolded delta functions \eqref{T2ZNdelta}
that appear in these graphs by one, and written that one out
explicitly. In addition, we have performed a change of coordinates  
$z_1 \ra e^{-\frac i2b\gf}\, z_1$ and symmetrized the result explicitly
under $b \ra \sm b$, by defining 
\equ{
\cP_{b}(X_2, X_1)
~=~
\frac 1{(\Box_6-m^2)_2} 
\gd^6\big(z_2-e^{\sm \frac i2 b\gf}\, z_1\big) \, 
\frac 1{(\Box_6-m^2)_2} 
\gd^6\big(z_2- e^{\frac i2 b\gf}\, z_1\big)
~, 
\labl{PropcP}
}
which satisfies: $\cP_{\sm b} = \cP_{b}$. 
Here we have introduced an IR regulator mass $m$ to identify the
quadratic divergences in the dimensional reduction (DR) scheme. 
(In 5D $m$ denotes the mass of the hyper multiplet.) 
In the delta functions we have only indicated the compact 
coordinates explicitly, as only there one encounters the 
phase $\exp({\pm \frac i2 b\gf})$. We can read off from
\eqref{1LoopAmpl} whether the combination of self--energy 
diagrams of figure \ref{fg:VV_VS_bSS} has localized contributions. 
The contribution $b=0$ gives the bulk amplitude. The 
contributions $b\neq0$, sourced by the fixed points, depend on
the orbifold:  
\items{
\item For the $\Intr_2$ orbifold and the $\Intr_2$ sector ($b=N/2$) of 
even ordered $\Intr_N$ orbifolds we find no localized contributions,
independently of the hyper multiplet twist eigenvalue $a_+$, since
$\cos(a_++\frac 12) \gp \!=\! 0$. 
\item However, for a generic $\Intr_N$ orbifold with $N > 2$ we find
contributions sourced by the fixed points for the sectors 
$b = \pm 1, \ldots,$ $\pm[(N-1)/2]$.
}
This confirms and extends the results of section \ref{sc:S1} based on
a component analysis on $S^1/\Intr_2$.

\begin{figure}
\equ{
\arry{l}{
\gS_{VV} ~=~ 
\raisebox{-1ex}{\scalebox{0.35}{\mbox{\input{VV_I.pstex_t}}}}
~ + ~
\raisebox{-2ex}{\scalebox{0.35}{\mbox{\input{VV_A.pstex_t}}}}
~+~
\raisebox{-2ex}{\scalebox{0.35}{\mbox{\input{VV_B.pstex_t}}}}
\\[3ex]
\gS_{VS} ~=~ 
\raisebox{-2ex}{\scalebox{0.35}{\mbox{\input{VS.pstex_t}}}}
\hfill 
\gS_{\bS S} ~=~ 
\raisebox{-2ex}{\scalebox{0.35}{\mbox{\input{bSS.pstex_t}}}}
}
\non}
\caption{The gauge self--energy supergraphs 
are drawn. The wavy and straight lines indicate the superfields  
$V$, $S$ and $\bS$. The lines with double arrows
depict the hyper multiplet propagators \eqref{Props}.}
\labl{fg:VV_VS_bSS}
\end{figure}

\subsection{Higher derivative counter terms and renormalized gauge
couplings} 
\labl{sc:counter}

After having distinguished bulk and localized fixed point
contributions, we determine the counter terms required by this 
theory. The bulk contribution, $b=0$, is proportional to the 6D
momentum integral  
\equ{
\int \frac {\d^D P}{(2\gp)^D} \gD_{P\, K}^m ~=~ 
\int \frac{\d^d p}{(2\gp)^d} \, \frac 1{\text{Vol}_W} \sum_{n\in \gL_K} 
\frac 1{p^2 + |n|^2 + m^2} \, 
\frac 1{(p-k)^2 + |n-l|^2 + m^2}~. 
\labl{DivInt6D}
} 
The sum is over the 2D KK lattice $\gL_K$, see
table \ref{tb:VolLatt}. The dimensionally regularized $D=2+d=6-2 \ge$
integral is defined to include the factor $1/\gm^{d-4}$ so as to keep
the mass dimension canonical throughout the regularization process.  
In the Appendix some steps are given to show that
\eqref{DivInt6D} can be represented as  
\equ{
\int \frac {\d^D P}{(2\gp)^D} \gD_{P\, K}^m ~=~ 
\frac {\gm^{2}}{(4 \gp)^{\frac {D}{2}}}
\int
_0^1 \d s \int
_0^\infty \frac{\d t}{t^{\frac d2}}\ 
e^{\dsp- t \big\{ s(1-s) (k^2+|l|^2) + m^2\big\}/\gm^2}\ 
\gth_W\brkt{{0}}{\sm sl}\big(\sfrac {i \gm^2}{2 t} \big)~.
\labl{ExprW}
}
The Jacobi theta function $\gth_W\brkt{0}{\sm sl}$ associated
with the winding mode lattice $\gL_W$ (defined in \eqref{Thetas}) is
obtained after a Poisson resummation.

This expression contains a lot of information: From the expression of
$\gth_W\brkt{0}{\sm sl}$ given in \eqref{Thetas}, it follows that 
$\smash{\gth_W\brkt{0}{\sm sl} \ra 1}$ 
in the UV ($t \ra 0$), since all terms  
in the winding mode sum are exponentially suppressed. 
Therefore, to determine the counter terms we can put
$\gth_W\brkt{0}{\sm sl}$ equal to $1$. This shows that the bulk
counter terms respect the 6D Lorentz invariance, since the external
momenta appear in the combination $K^2 = k^2 + |l|^2$ only. 
The difference $\gth_W\brkt{0}{\sm sl} - 1$ encodes the threshold
corrections due to the (Poisson resummed) KK modes. 
Such threshold corrections have been studied for external 
zero modes ($l =0$) in the effective field theory limit of string theory
\cite{Kaplunovsky:1987rp,Dixon:1990pc} and extra dimension 
models \cite{Ghilencea:2002ff}. Our result shows that for non--zero
mode KK states the threshold corrections will be
different from those for the zero modes.  (This is related to non--local
corrections to KK masses studied in ref.\ \cite{Cheng:2002iz}.) 
The counter terms are determined by $I_2^{\rm div}$ given in
\eqref{DivParts} of the Appendix. The divergence proportional to 
$K^2$ in \eqref{DivParts} requires the higher derivative 
counter term with a dimensionless coupling $1/h^2$: 
\equa{ 
 \cL_{\rm gauge}^{\rm hd}
~=~& - \frac 1{2h^2N}
\int \d^2 \gth\, W^\ga \,\Box_6\, W_\ga ~+~
\labl{LgaugeHD}
\\[1ex] 
&-~\frac 1{h^2N} 
\int \d^4 \gth\, \Big(
\der V \,\Box_6\, \bder V +  \bS \,\Box_6\, S  
- \sqrt 2\, \bder V  \,\Box_6\,S  - \sqrt 2\, \der V  \,\Box_6\, \bS 
\Big)~. 
\non 
}

To conclude the Letter we compute the renormalized gauge couplings. 
Here we only give the parts of the couplings which do not dependent on
the external KK momenta. In addition we neglect the
finite threshold correction due to the resummed KK states. 
In a complete treatment the brane localized kinetic terms should be
taken into account \cite{Carena:2002me}.
For the sake of brevity we ignore all these complications; in a future
publication we return to them in detail \cite{GNH}. 
The renormalizations of bulk gauge couplings $g$ and $h$, defined in
\eqref{Lgauge} and \eqref{LgaugeHD} respectively,  are given by  
\equ{
\frac 1{g_R^2} ~=~ \frac 1{g^2} + \frac {2q^2}{(4\gp)^3} 
\, m^2\, \Big[ 1 + \ln \Big( \frac{\gm^2}{m^2}\Big) \Big]~, 
\qquad 
\frac 1{h^2_R} ~=~ \frac 1{h^2} - \frac 16\,  \frac {2q^2}{(4\gp)^3}
\, \ln \Big( \frac{\gm^2}{m^2}\Big)~,
\labl{RenormBulk}
}
in the $\overline{\rm DR}$ scheme. (The 5D result is discussed in
section \ref{sc:HyperS1}.) The coupling $h$ renormalizes as
anticipated by \cite{Smilga:2004cy}.

The localized contributions with $b \neq 0$ can be analyzed in a
similar fashion. Neither of the KK loop momenta $n_1, n_2$ are free
since they are fixed by the external KK momenta $l_1, l_2$ as 
\equ{
\pmtrx{ n_1 \\ n_2}  ~=~ 
\frac {\sm i}{2 \sin \frac 12 b \gf} \, 
\pmtrx{1 & \sm e^{\sm\frac i2 b \gf } \\ \sm 1 & e^{\frac i2 b \gf} }
\, 
\pmtrx{ ~l_1 \\ \sm l_2 }~. 
}
(This generalizes the violation of the KK--momenta that we 
encountered in section \ref{sc:fermionS1}.)  Therefore the divergences
can only come from the 4D momentum $p$ in the loop: 
\equ{
\int \frac{\d^d p}{(2\gp)^d} \, 
\frac 1{p^2 + m^2 + |n_1|^2 } \,  
\frac 1{(p-k)^2 + m^2+|n_2|^2}~, 
} 
The divergent part $I_0^{\rm div}$ (given in \eqref{DivParts}) of this
integral is independent of  the external KK numbers $l_1$ and $l_2$ up to 
finite renormalizations which are ignored here. The running of the
fixed point gauge couplings $g_b$, given in \eqref{LgaugeFix}, reads:
\equ{
\frac 1{(g^2_b)_R} ~=~ \frac 1{g_b^2} 
- \frac {2q^2}{(4\gp)^2}\, 
\cos \big(a_+ + \sfrac 12 \big)b \gf\, \cos \sfrac 12 b \gf \, 
\ln \Big( \frac{\gm^2}{m^2}  \Big)~. 
\labl{RenormFix}
}

Finally, we note that in the limit where we take the IR regulator $m$
to zero, $h_R$ and $(g_b)_R$ suffer from logarithmic IR singularities,
and the coupling $g_R$ becomes equal to its tree level value. All
these statements of course ignore important finite volume effects that
lead to finite KK number dependent renormalizations and will have to be
discussed in \cite{GNH}.

\appendix 
\def\theequation{A.\arabic{equation}} 
\setcounter{equation}{0}

\section*{Appendix: Regularization of the common scalar integrals}

We extract the divergent parts of the integral \eqref{DivInt6D} in 4,
5, and 6D.  Using a Schwinger proper time reparameterization $t$ and
a Feynman parameter $s$ this integral can be expressed as 
\equ{
\int \frac {\d^D P}{(2\gp)^D} \gD_{P\, K}^m =
\frac 1{(4\gp)^{\frac d2} \, \text{Vol}_W}  
\int
_0^1 \d s \int
_0^\infty \frac{\d t}{t^{\frac d2-1}}\ 
e^{\dsp- t \big\{ s(1-s) (k^2+|l|^2) + m^2\big\}/\gm^2}\ 
\gth_K\brkt{{sl}}{0}\big(\sfrac {2i t}{\gm^2} \big)~.
\labl{ExprK}
}
Here we have introduced the Jacobi theta functions for the
KK and winding mode lattices 
\equ{
\gth_K\brkt{\ga}{\gb}(\gt) ~= \sum_{n \in \gL_K} 
e^{i \frac \gt2 |n-\ga|^2 - i (\bn - \bga) \bgb - i (n-\ga) \gb}~, 
\quad 
\gth_W\brkt{\gb}{\ga}(\gt) ~= \sum_{w \in \gL_W} 
e^{2 i \gt |w-\gb|^2 - i (\bw - \bgb) \bga - i (w-\gb) \ga}~, 
\labl{Thetas}
} 
see table \ref{tb:VolLatt}. 
They are related to each other via a Poisson resummation: 
\equ{
\gth_K\brkt{\ga}{\gb}(\gt) ~=~ 
 \big(\sfrac{2\gp}{\sm i\, \gt}\big)^{\frac{D-d}2} 
\, \text{Vol}\,
\gth_W\brkt{\gb}{\sm \ga}\big(\sfrac{\sm 1}{\gt}\big)~.  
}
Applying this relation to \eqref{ExprK} we obtain the formula given in
\eqref{ExprW} given in the main text. 
To determine the divergent parts of \eqref{ExprW}, we define the
integral expression  
\equ{
I_{D\sm d}(K^2, m^2) ~=~ 
\frac {\gm^{D-d}}{(4 \gp)^{\frac {D}{2}}}
\int
_0^1 \d s \int
_0^\infty \frac{\d t}{t^{\frac D2-1}}\ 
e^{\dsp- t \big\{ s(1-s) K^2 + m^2 \big\}/\gm^2}~.
\labl{Iint} 
}
Provided that $d \in \Cplx$ is suitably chosen, this expression is
convergent and can be cast into the form  
\equ{
I_{D\sm d} ~=~ 
\frac 1{(4\gp)^2} \Big( \frac {m^2}{4\gp}  \Big)^{\frac{D-d}2}
\Big( 4\gp\, \frac{\gm^2}{m^2} \Big)^{2 - \frac d2}
\sum_{n \geq 0} (-)^n \, 
\frac {\gG(n+2 - \frac D2) n!}{(2n+1)!} \, 
\Big( \frac{K^2}{m^2} \Big)^{n}
~. 
}
The terms with $0 \leq n \leq \frac D2 - 2$ correspond to the terms in
the Taylor expansion of \eqref{Iint} in $K^2$ with divergent
coefficients, if we had not analytically continued $D \in \Cplx$. We
refer to these terms by the notation $I^{\rm div}_{D\sm d}(K^2,m^2)$. 
Explicitly, we have in $D\sm d = 0, 1, 2$ extra dimensions and with 
$d = 4 - 2 \ge$: 
\equ{
\arry{l}{\dsp 
I_{0}^{\rm div} ~=~ \frac 1{(4\gp)^2} 
\Big( \frac 1\ge - \gg + \ln\big( 4\gp \sfrac{\gm^2}{m^2}\big) \Big)~, 
\qquad \qquad 
I_{1}^{\rm div} ~=~ - \frac 1{(4\gp)^2} {|m|}~, 
\\[2ex] \dsp 
I_2^{\rm div} ~=~ 
- \frac 1{(4\gp)^3} \Big[ m^2 + 
\Big(  \frac 1\ge - \gg + \ln \big( 4\gp \sfrac{\gm^2}{m^2}\big) \Big)
\Big( {m^2} + \frac 16\, {K^2} \Big)
\Big]~, 
}
\labl{DivParts} 
}
where $\gg$ is the Euler constant. The case $D\sm d=0$
gives the familiar 4D expression.

\bibliographystyle{letter}
{\small
\bibliography{paper}
}

\newpage

\end{document}